\documentclass{article}
\usepackage{amsmath}
\usepackage{graphicx}

\input{tcilatex}
\begin{document}

\title{Quantum State Tomography and Quantum Games }
\author{Ahmad Nawaz\thanks{%
Email: ahmad@ele.qau.edu.pk} \\
National Centre for Physics, Quaid-i-Azam University Campus,\\
Islamabad, Pakistan}
\maketitle

\begin{abstract}
We develop a technique for single qubit quantum state tomography using the
mathematical setup of generalized quantization scheme for games. In this
technique Alice sends an unknown pure quantum state to Bob who appends it
with $\left\vert 0\right\rangle \left\langle 0\right\vert $ and then applies
the unitary operators on the appended quantum state and finds the payoffs
for Alice and himself.\ Is is shown that for a particular set of unitary
operators these payoffs are equal to Stokes parameters for an unknown
quantum state. In this way an unknown quantum state can be measured and
reconstructed. Strictly speaking this technique is not a game as no
strategic competitions are involved.
\end{abstract}

\section{Introduction}

All information about a quantum system is encoded in the state of the system
but it is one of the great challenges for experimentalists to measure the
state of the quantum system perfectly \cite{schleich}. This is due to the
fact that the state of a quantum system is not an observable in quantum
mechanics \cite{peres} that makes it impossible to perform all measurements
on the single state to extract the whole information about the system. On
the other hand no-cloning theorem does not allow to create a perfect copy of
the system without prior knowledge about its state \cite{wootters}. Hence,
there remains no way, even in principle, to infer the unknown quantum state
of a single quantum system \cite{ariano}. However it is possible to estimate
the unknown quantum state of a system when many identical copies of the
system are available. This procedure of reconstructing an unknown quantum
state through a series of measurements on a number of identical copies of
the system is called quantum state tomography. In this process each
measurement gives a new dimension of the system and therefore, infinite
number of copies are required to reconstruct the exact state of a quantum
system.\textrm{\ }The problem of quantum state tomography was first
addressed by Fano \cite{fano} who recognized the need to measure two non
commuting observables. However it remained mere speculation until original
proposal for quantum tomography and its experimental verification \cite%
{ariano,vegal,raymer}. Since than it is being\ applied for the measurement
of photon statistics of a semiconductor laser \cite{munroe}, reconstruction
of density matrix of squeezed vacuum \cite{schiller} and probing the
entangled states of light and ions \cite{paris}\textrm{. The other main
tasks where the role of quantum state tomography is necessary for the
complete characterization of the state of the system are: to study the
effects of decoherence \cite{white}, to optimize the performance of quantum
gates \cite{brien}, to quantify the amount of information that various
parties can obtain by quantum communication protocols \cite{langford} and
utilization of quantum error correction protocols in real world situations
effectively \cite{altepeter}.}

In this paper by making use of the mathematical setup of generalized
quantization scheme for games \cite{nawaz} a technique for quantum state
tomography is developed.\textrm{\ Strictly speaking this arrangement is not
a game but only the mathematical setup of quantum games is used as a tool} \ 
\textrm{It works as follows: Alice sends an unknown pure quantum state }$%
\rho $\textrm{\ to Bob who appends it with }$\left\vert 0\right\rangle
\left\langle 0\right\vert $\textrm{\ resulting the initial state of the game
as }$\rho _{in}=\left\vert 0\right\rangle \left\langle 0\right\vert \otimes
\rho .$\textrm{\ On this appended quantum state Bob applies unitary operator 
}$U=U_{A}\otimes U_{B}$\textrm{\ and finds the the payoffs }$\left(
\$_{A},\$_{B}\right) $ using the \textrm{predefined payoff operators }$P^{A}$%
\textrm{\ and }$P^{B}$\textrm{.\ For a particular set of unitary operators
(strategies) and payoff operators these payoffs become Stokes parameters of
the given quantum state }$\rho $\textrm{. In this way an unknown quantum
state can be measured and reconstructed. It is common to use entangled state
as an input for quantum games when one is interested to find the solution of
a game such as the resolution of dilemma in prisoner dilemma game \cite%
{eisert}. But the payoffs cannot be independent of initial quantum state
whether the state is product or entangled and hence the information about
the input quantum state is reflected as a function of payoffs at the output.
This makes it possible to estimate an unknown quantum state. Furthermore
this technique does not improve the standard technique of quantum state
tomography but }it is a step forward for strengthening the established link
between quantum games and quantum information theory \cite{meyer}.

This paper is arranged as follows:- In section (\ref{stokes}) we present a
brief introduction to single qubit tomography following Refs. \cite%
{altepeter,chuang}, in section (\ref{State Tomography}) we present our
technique for quantum state tomography and section (\ref{conclusion})
concludes the results.

\section{\label{stokes}The Stokes Parameters Representation of Qubit}

Any single qubit density matrix $\rho $ can uniquely be represented with the
help of three parameters $\left\{ S_{1},S_{2},S_{3}\right\} $ and Pauli
matrices $\sigma _{i}^{\prime }s$\ by the expression 
\begin{equation}
\rho =\frac{1}{2}\underset{i=0}{\overset{3}{\tsum }}S_{i}\sigma _{i},
\label{stokes representation}
\end{equation}%
where $S_{0}=1$ and the other parameters obey the relation $\underset{i=0}{%
\overset{3}{\tsum }}S_{i}^{2}\leq 1$. The parameters, $S_{i}$\ are called
Stokes parameters and for a quantum state $\rho $\ these can be calculated
as 
\begin{equation}
S_{i}=\text{Tr}\left( \sigma _{i}\rho \right) .
\end{equation}%
Physically these parameters give the outcome of a projective measurements as

\begin{eqnarray}
S_{0} &=&P_{\left| 0\right\rangle }+P_{\left| 1\right\rangle }  \notag \\
S_{1} &=&P_{\frac{1}{\sqrt{2}}\left( \left| 0\right\rangle +\left|
1\right\rangle \right) }-P_{\frac{1}{\sqrt{2}}\left( \left| 0\right\rangle
-\left| 1\right\rangle \right) }  \notag \\
S_{2} &=&P_{\frac{1}{\sqrt{2}}\left( \left| 0\right\rangle +i\left|
1\right\rangle \right) }-P_{\frac{1}{\sqrt{2}}\left( \left| 0\right\rangle
-i\left| 1\right\rangle \right) }  \notag \\
S_{3} &=&P_{\left| 0\right\rangle }-P_{\left| 1\right\rangle }
\end{eqnarray}%
where $P_{\left| i\right\rangle }$ is the probability to measure state $%
\left| i\right\rangle $ given by 
\begin{eqnarray}
P_{\left| i\right\rangle } &=&\left\langle i\right| \rho \left|
i\right\rangle  \notag \\
&=&\text{Tr}\left( \left| i\right\rangle \left\langle i\right| \rho \right) .
\end{eqnarray}%
If we are provided with many copies of a quantum state then with the help of
orthogonal set of matrices $\frac{\sigma _{0}}{\sqrt{2}},\frac{\sigma _{1}}{%
\sqrt{2}},\frac{\sigma _{2}}{\sqrt{2}},\frac{\sigma _{3}}{\sqrt{2}}$ the
density matrix (\ref{stokes representation}) can be written as 
\begin{equation}
\rho =\frac{\text{\textrm{Tr}}(\rho )\sigma _{0}+\text{\textrm{Tr}}(\rho
\sigma _{1})\sigma _{1}+\text{\textrm{Tr}}(\rho \sigma _{2})\sigma _{2}+%
\text{\textrm{Tr}}(\rho \sigma _{3})\sigma _{3}}{2}.  \label{rho}
\end{equation}%
\textrm{where the} expression like \textrm{Tr}$(\rho \sigma _{i})$\
represents the expectation value of the observable. To estimate \textrm{Tr}$%
(\rho \sigma _{3}),$ for example, we measure $\sigma _{3}$ for $m$ numbers
of time giving the values $z_{1},z_{2,}.....,z_{m}$ all equal to +1 or -1.
The average $\tsum \frac{z_{i}}{m_{i}}$ is an estimate to true value of the
quantity \textrm{Tr}$(\rho \sigma _{3}).$ By central limit theorem this
estimate has standard deviation $\frac{\Delta \sigma _{3}}{m}$ where $\Delta
\sigma _{3}$\ is the standard deviation for single measurement of $\sigma
_{3}$ that is upper bounded by 1. Therefore, the standard deviation for
estimate $\tsum \frac{z_{i}}{m_{i}}$ is at most $\frac{i}{\sqrt{m}}.$The
standard deviation for each of the measurement in Eq. (\ref{rho}) is the
same \cite{chuang}. In this way with the help of Eq. (\ref{rho}) tomography
can be performed for an unknown single qubit state.

\subsection{Single Qubit Tomography}

A single qubit state can very conveniently be represented by a vector in
three dimensional vector space spanned by Pauli matrices. This
representation provides very helpful way for geometrical visualization of
single qubit state, where all the legal states fall within a unit sphere
(Bloch sphere). In this representation all the pure states lie on the
surface of the sphere and mixed states fall inside the sphere. The pure
states can be written as 
\begin{equation}
\left| \psi \right\rangle =\cos \frac{\theta }{2}\left| 0\right\rangle
+e^{i\phi }\sin \frac{\theta }{2}\left| 1\right\rangle  \label{initial state}
\end{equation}%
where\textrm{\ }$\theta $\textrm{\ }and\textrm{\ }$\phi $\textrm{\ }map them
on the surface of the sphere. Any state $\left| \psi \right\rangle $ and its
orthogonal component $\left| \psi ^{\perp }\right\rangle $ fall on two
opposite points on the surface of the sphere such that the line connecting
these points form the axis of the sphere.

\begin{figure}[th]
\centering\includegraphics[scale=1]{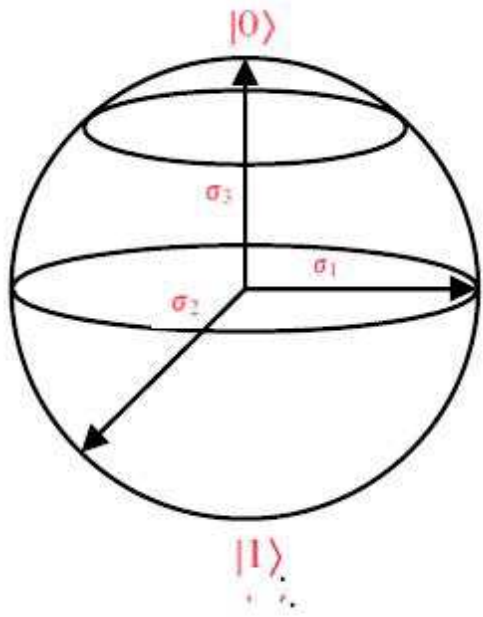}
\caption{The measuremsnt in $\protect\sigma _{3}$ basis confines the unkown
quantum state to a plane $z=\cos \protect\theta .$}
\label{tomography1}
\end{figure}

For the tomography of an unknown single qubit state three consecutive
measurements are required. Each measurement gives one dimension of the
system until one becomes aware of all dimensions after the complete set of
measurement. For example, a single qubit state $\rho =\left\vert \psi
\right\rangle \left\langle \psi \right\vert $ where $\left\vert \psi
\right\rangle $ is defined in Eq. (\ref{initial state}), can be expressed as 
\begin{equation}
\rho =\frac{1}{2}\left( \sigma _{0}+\sin \theta \cos \phi \text{ }\sigma
_{1}+\sin \theta \sin \phi \text{ }\sigma _{2}+\cos \theta \text{ }\sigma
_{3}\right)  \label{Stokes}
\end{equation}%
Comparing Eqs. (\ref{stokes representation}) and (\ref{Stokes}) the Stokes
parameters for this state become 
\begin{equation}
S_{1}=\sin \theta \cos \phi ,\text{ }S_{2}=\sin \theta \sin \phi ,\text{ }%
S_{3}=\cos \theta .  \label{stokes parameters}
\end{equation}%
For an unknown state of the form Eq. (\ref{Stokes}) when a measurement is
performed in $\sigma _{3}$ basis it confines the state to a plane $z=\cos
\theta $; as shown in Fig. (\ref{tomography1}).

\begin{figure}[th]
\centering\includegraphics[scale=1]{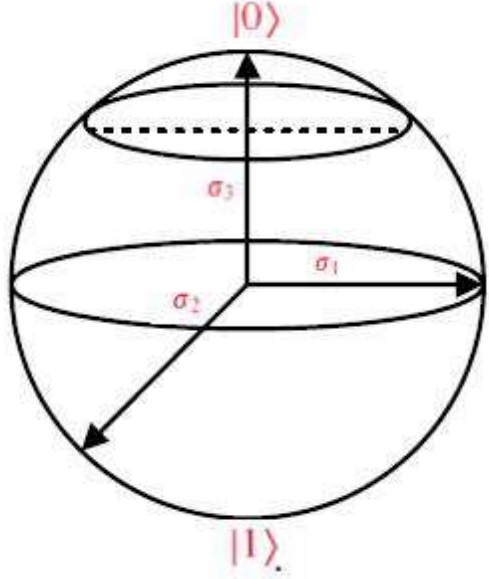}
\caption{The measurement in $\protect\sigma _{2}$ basis confines the state
to $y=\sin \protect\theta \sin \protect\phi $ plane. When this measurement
is combined with first measurement the unknown state reduces to a line
parrallel to x-axis.}
\label{tomography-2}
\end{figure}

Then a measurement in $\sigma _{2}$\ basis is performed that further
confines it to the plane $y=\sin \theta \sin \phi $. The combined effect of
both these measurements restricts the unknown quantum state to a line
parallel to x-axis as shown in Fig. (\ref{tomography-2}).

At last the measurement in $\sigma _{1}$ basis pinpoints the state as point
lying on this line (resulting from the intersection of $y$ and $z$ planes)
at distance $x=\sin \theta \cos \phi $; as illustrated in Fig. (3). Since
the resultant state is due the intersection of three orthogonal planes
therefore the order of these measurements is immaterial in the whole process.

\begin{figure}[th]
\centering\includegraphics[scale=1]{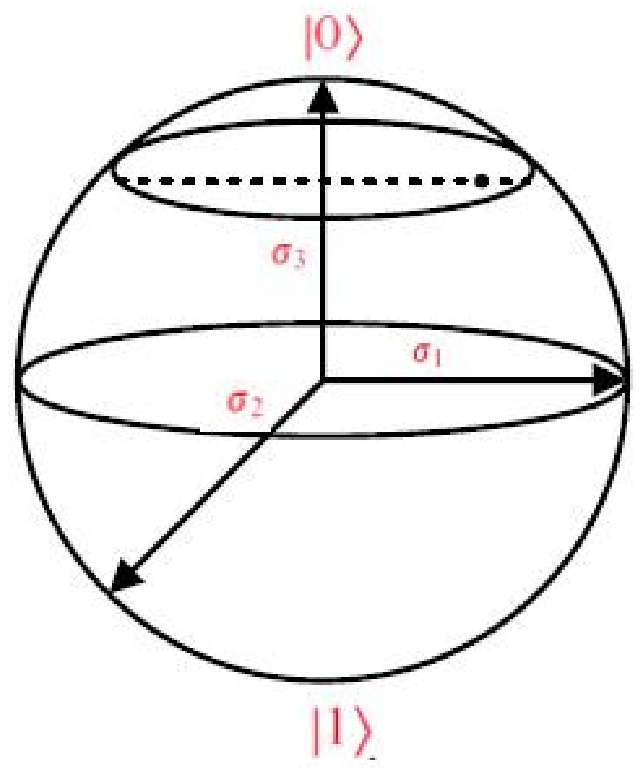}
\caption{The last measurement in $\protect\sigma _{1}$ basis pinpoints the
state as point \ that results from the intersection of three orthogonal
planes.}
\label{tomography}
\end{figure}

In the next section we show that how a single qubit quantum state tomography
can be performed using mathematical setup of generalized quantization scheme
for games \cite{nawaz}.

\section{\label{State Tomography}Quantum State Tomography by Mathematical
Setup of Quantum Games}

Let Alice forwards an unknown quantum state of the form of Eq. (\ref{initial
state}) to Bob who appends it with $\left\vert 0\right\rangle \left\langle
0\right\vert $ resulting the initial state of the game as 
\begin{equation}
\rho _{in}=\cos ^{2}\frac{\theta }{2}\left\vert 00\right\rangle \left\langle
00\right\vert +\sin ^{2}\frac{\theta }{2}\left\vert 01\right\rangle
\left\langle 01\right\vert +\frac{\sin \theta e^{i\phi }}{2}\left(
\left\vert 01\right\rangle \left\langle 00\right\vert +e^{-i2\phi
}\left\vert 00\right\rangle \left\langle 01\right\vert \right)
\label{appended}
\end{equation}%
and then applies the unitary operator 
\begin{equation}
U=U_{A}\otimes U_{B}  \label{unitary operator}
\end{equation}%
on the appended state (\ref{appended}). Where for$\ k=A$, $B$ we have
defined 
\begin{equation}
U_{k}=\cos \frac{\beta _{k}}{2}R_{k}+\sin \frac{\beta _{k}}{2}P_{k}
\end{equation}%
with $0\leq \beta _{k}\leq \pi .$ The operations of $R_{k}$ and $P_{k}$\emph{%
\ on} $\left\vert 0\right\rangle $ and $\left\vert 1\right\rangle $ are
defined as

\begin{align}
R_{k}\left\vert 0\right\rangle & =e^{i\alpha _{k}}\left\vert 0\right\rangle
\ ,\text{ \ \ }R_{k}\left\vert 1\right\rangle =e^{-i\alpha _{k}}\left\vert
1\right\rangle ,  \notag \\
P_{k}\left\vert 0\right\rangle & =-\left\vert 1\right\rangle ,\text{ \ \ \ \
\ }P_{k}\left\vert 1\right\rangle =\left\vert 0\right\rangle .
\label{operators}
\end{align}%
\textrm{The unitary operators (}\ref{unitary operator})\textrm{\ transform
the initial state (\ref{appended}) to} 
\begin{equation}
\rho _{f}=\left( U_{A}\otimes U_{B}\right) \rho _{in}\left( U_{A}\otimes
U_{B}\right) ^{\dagger }.  \label{final-state}
\end{equation}%
and then Bob finds the payoffs by using the formula 
\begin{equation}
\$^{k}(U_{A},U_{B},\theta ,\phi )=\text{Tr}(P^{k}\rho _{f})\text{,}
\label{payoff formula}
\end{equation}%
where $P^{k}$ the payoff operators given as

\begin{equation}
P^{k}=\$_{00}^{k}\left\vert 00\right\rangle \left\langle 00\right\vert
+\$_{01}^{k}\left\vert 01\right\rangle \left\langle 01\right\vert
+\$_{10}^{k}\left\vert 10\right\rangle \left\langle 10\right\vert
+\$_{11}^{k}\left\vert 11\right\rangle \left\langle 11\right\vert ,
\label{payoff operator}
\end{equation}%
with $\$_{ij}^{k}$ as the entries of payoff matrix in $ith$ row and $jth$
column for player $k$. With the help of Eqs. (\ref{initial state}, \ref%
{payoff operator}, \ref{payoff formula}) the payoffs come out to be 
\begin{align}
\$^{k}(U_{A},U_{B},\theta ,\phi )& =\left( \$_{00}^{k}\chi
+\$_{11}^{k}\Omega +\$_{01}^{k}\xi +\$_{10}^{k}\eta \right) \cos ^{2}\frac{%
\theta }{2}+\left( \$_{00}^{k}\xi +\$_{11}^{k}\eta +\$_{01}^{k}\chi \right. +
\notag \\
& \left. \$_{10}^{k}\Omega \right) \sin ^{2}\frac{\theta }{2}+\left[ \left\{
\left( \$_{00}^{k}-\$_{01}^{k}\right) \Phi +\left(
\$_{10}^{k}-\$_{11}^{k}\right) \Theta \right\} \cos \alpha _{B}\right] \sin
\theta \cos \phi +  \notag \\
& \left[ \left\{ \left( \$_{00}^{k}-\$_{01}^{k}\right) \beta +\left(
\$_{10}^{k}-\$_{11}^{k}\right) \Theta \right\} \sin \alpha _{B}\right] \sin
\theta \sin \phi ,  \label{payoff}
\end{align}%
where 
\begin{eqnarray}
\chi &=&\cos ^{2}\frac{\beta _{A}}{2}\cos ^{2}\frac{\beta _{B}}{2},\text{ \ }%
\xi =\cos ^{2}\frac{\beta _{A}}{2}\sin ^{2}\frac{\beta _{B}}{2},  \notag \\
\Omega &=&\sin ^{2}\frac{\beta _{A}}{2}\sin ^{2}\frac{\beta _{B}}{2},\text{
\ }\eta =\sin ^{2}\frac{\beta _{A}}{2}\cos ^{2}\frac{\beta _{B}}{2},  \notag
\\
\Phi &=&\frac{1}{2}\cos ^{2}\frac{\beta _{A}}{2}\sin \beta _{B},\text{ \ }%
\Theta =\frac{1}{2}\sin ^{2}\frac{\beta _{A}}{2}\sin \beta _{B}.
\end{eqnarray}%
It is evident from Eq. (\ref{payoff}) that the payoffs contains the
information about the initial quantum state in terms of Stokes parameters
defined in Eq. (\ref{stokes parameters}) that can be extracted by using a
suitable set of strategies. For $\$_{00}^{A}=\$_{10}^{A}=\$_{01}^{B}=%
\$_{11}^{B}=1,\$_{11}^{A}=\$_{01}^{A}=\$_{00}^{B}=\$_{10}^{B}=-1$ Bob
performs the following steps for single qubit quantum state tomography

Step (1) When $\beta _{A}=\beta _{B}=\alpha _{B}=\frac{\pi }{2}$ with the
help of Eq. (\ref{payoff}) we get 
\begin{eqnarray}
\$^{A} &=&\sin \theta \sin \phi ,  \notag \\
\$^{B} &=&-\sin \theta \sin \phi .  \label{case-1}
\end{eqnarray}%
Comparing the result (\ref{case-1}) with Eq. (\ref{stokes parameters}) we
see that the payoff of Alice is one of the Stokes parameters.

Step (2) When $\beta _{A}=\beta _{B}=\frac{\pi }{2}$ and $\alpha _{2}=0$
then Eq. (\ref{payoff}) reduces to 
\begin{eqnarray}
\$^{A} &=&\sin \theta \cos \phi ,  \notag \\
\$^{B} &=&-\sin \theta \cos \phi .  \label{case-2}
\end{eqnarray}%
Comparing Eqs. (\ref{case-2})\ and (\ref{stokes parameters}) it is evident
that it is also one the Stokes parameters.

Step (3) When $\beta _{A}=\beta _{B}=0$ then Eq. (\ref{payoff}) gives 
\begin{eqnarray}
\$^{A} &=&\cos \theta ,  \notag \\
\$^{B} &=&-\cos \theta .  \label{case-3}
\end{eqnarray}%
Comparison of the result (\ref{case-3}) with Eq. (\ref{stokes parameters})
shows the payoff of Alice is third Stokes parameter.

From Eqs. (\ref{case-1}) (\ref{case-2}) and (\ref{case-3}) we see that the
payoffs are equal to the Stokes parameters of quantum state that helps us to
reconstruct the quantum state. Furthermore the standard deviation for all of
the above cases is bounded above by 1\textrm{.} Furthermore this technique
is simple and not beyond the reach of recent technology \cite{du-2,zhou}.

\section{\label{conclusion}Conclusion}

The state of the quantum system contains all the information about the
system. In classical mechanics it is possible in principle, to devise a set
of measurements that can fully recover the state of the system. In quantum
mechanics two fundamental theorems, Heisenberg uncertainty principle and no
cloning theorem forbid to recover the state of a quantum system without
having some prior knowledge. This problem, however, can be solved with the
help of quantum state tomography. Where an unknown quantum state is
estimated through a series of measurements on a number of identical copies
of a system. Here we showed that how an unknown quantum state can be
reconstructed by making use of mathematical setup of generalized
quantization scheme of games. In our technique Alice sends an unknown pure
quantum state to Bob who appends it with $\left\vert 0\right\rangle
\left\langle 0\right\vert $ and then applies the unitary operators on the
appended quantum state and finds the payoffs for Alice and Bob.\ It is shown
that for a particular set of unitary operators the payoffs become equal to
Stokes parameters for the unknown quantum state. In this way an unknown
quantum state can be measured and reconstructed.

\end{document}